\documentclass[final,3p,times,twocolumn]{elsarticle}
\usepackage{xcolor,etoolbox}
\usepackage{ulem}
\usepackage{amssymb}
\usepackage{amsthm}
\usepackage{amsmath}
\usepackage{here}

\journal{Physics Letters B}

\begin{document}

\begin{frontmatter}

%% Title, authors and addresses

%% use the tnoteref command within \title for footnotes;
%% use the tnotetext command for theassociated footnote;
%% use the fnref command within \author or \address for footnotes;
%% use the fntext command for theassociated footnote;
%% use the corref command within \author for corresponding author footnotes;
%% use the cortext command for theassociated footnote;
%% use the ead command for the email address,
%% and the form \ead[url] for the home page:
%\title{Title\tnoteref{label1}}

%%%%% Author list %%%%%

\affiliation[tohoku]{Research Center for Neutrino Science, Tohoku University, Sendai 980-8578, Japan}
\affiliation[obihiro]{Department of Human Science, Obihiro University of Agriculture and Veterinary Medicine, Obihiro 080-8555, Japan}
\affiliation[ipmu]{Kavli Institute for the Physics and Mathematics of the Universe (WPI), The University of Tokyo Institutes for Advanced Study, The University of Tokyo, Kashiwa, Chiba, 277-8583, Japan}
\affiliation[GPPU]{Graduate Program on Physics for the Universe, Tohoku University, Sendai 980-8578, Japan}
\affiliation[gakusai]{Frontier Research Institute for Interdisciplinary Sciences, Tohoku University, Sendai 980-8578, Japan}
\affiliation[osaka]{Graduate School of Science, Osaka University, Toyonaka, Osaka 560-0043, Japan}
\affiliation[rcnp]{Research Center for Nuclear Physics, Osaka University, Ibaraki, Osaka 567-0047, Japan}
\affiliation[tokushima]{Graduate School of Advanced Technology and Science, Tokushima University, Tokushima 770-8506, Japan}

\affiliation[rigaku]{Department of Physics, Tohoku University, Sendai 980-8578, Japan}

\affiliation[lbl]{Nuclear Science Division, Lawrence Berkeley National Laboratory, Berkeley, California 94720, USA}
\affiliation[hawaii]{Department of Physics and Astronomy, University of Hawaii at Manoa, Honolulu, Hawaii 96822, USA}
\affiliation[MIT]{Massachusetts Institute of Technology, Cambridge, Massachusetts 02139, USA}
\affiliation[ut]{Department of Physics and Astronomy, University of Tennessee, Knoxville, Tennessee 37996, USA}
\affiliation[tunl]{Triangle Universities Nuclear Laboratory, Durham, North Carolina 27708, USA and Physics Departments at Duke University, North Carolina Central University, and the University of North Carolina at Chapel Hill}
\affiliation[virginia]{Center for Neutrino Physics, Virginia Polytechnic Institute and State University, Blacksburg, Virginia 24061, USA}
\affiliation[washington]{Center for Experimental Nuclear Physics and Astrophysics, University of Washington, Seattle, Washington 98195, USA}
\affiliation[nikhef]{Nikhef and the University of Amsterdam, Science Park, 1098XG Amsterdam, Netherlands}
\affiliation[boston]{Boston University, Department of Physics, 590 Commonwealth Avenue, Boston, Massachusetts 02215, USA}

\fntext[icrrKamioka]{Present address: Kamioka Observatory, Institute for Cosmic-Ray Research, The University of Tokyo, Hida, Gifu 506-1205, Japan}
\fntext[riken]{Present address: Center for Advanced Photonics, RIKEN, Wako, Saitama, 351-0198, Japan}
\fntext[butsuryo]{Present address: Faculty of Health Sciences, Butsuryo College of Osaka, Sakai, Osaka 593-8328, Japan}
\fntext[takeuchi]{Present address: Faculty of Science, The University of Tokyo, Bunkyo-ku, Tokyo 113-0033, Japan}
\fntext[obara]{Present address: National Institutes for Quantum and Radiological Science and Technology (QST), Sendai 980-8579, Japan}
\fntext[kinken]{Present address: Institute for Materials Research, Tohoku University, 2-1-1 Katahira, Aoba-ku, Sendai, Miyagi, 980-8577, Japan}

\author[tohoku]{S.~Abe\fnref{icrrKamioka}}
\author[tohoku]{M.~Eizuka}
\author[tohoku]{S.~Futagi}
\author[tohoku]{A.~Gando}
\author[tohoku,obihiro]{Y.~Gando}
\author[tohoku]{S.~Goto}
\author[tohoku]{T.~Hachiya}
\author[tohoku]{K.~Hata\corref{cor1}}\ead{hata@awa.tohoku.ac.jp}
\cortext[cor1]{Corresponding author}
\author[tohoku]{K.~Hosokawa\fnref{icrrKamioka}}
\author[tohoku]{K.~Ichimura}
\author[tohoku]{S.~Ieki}
\author[tohoku]{H.~Ikeda}
\author[tohoku,ipmu]{K.~Inoue}
\author[tohoku]{K.~Ishidoshiro}
\author[tohoku]{Y.~Kamei\fnref{riken}}
\author[tohoku]{N.~Kawada}
\author[tohoku]{Y.~Kishimoto}
\author[tohoku,ipmu]{M.~Koga}
\author[tohoku]{M.~Kurasawa}
\author[tohoku]{T.~Mitsui\fnref{butsuryo}}
\author[tohoku]{H.~Miyake}
\author[tohoku]{D.~Morita}
\author[tohoku]{T.~Nakahata}
\author[tohoku]{R.~Nakajima}
\author[tohoku]{K.~Nakamura\fnref{butsuryo}}
\author[tohoku]{R.~Nakamura}
\author[tohoku]{R.~Nakamura}
\author[tohoku]{J.~Nakane}
\author[tohoku,GPPU]{H.~Ozaki}
\author[tohoku]{T.~Sakai} 
\author[tohoku]{I.~Shimizu}
\author[tohoku]{J.~Shirai}
\author[tohoku]{K.~Shiraishi}
\author[tohoku]{R.~Shoji}
\author[tohoku]{A.~Suzuki}
\author[tohoku]{A.~Takeuchi\fnref{takeuchi}}
\author[tohoku]{K.~Tamae}
\author[tohoku]{H.~Watanabe}
\author[tohoku]{K.~Watanabe}

% gakusai
\author[gakusai]{S.~Obara\fnref{obara}}

% IPMU
%\author[ipmu]{D.~Chernyak\fnref{dima}}
%\author[ipmu]{A.~Kozlov\fnref{sasha}}

% Osaka, RCNP
\author[osaka]{S.~Yoshida}
\author[rcnp]{S.~Umehara}
%\author[rcnp]{Y.~Takemoto\fnref{icrrKamioka}}

% Tokushima
\author[tokushima]{K.~Fushimi}
\author[tokushima]{K.~Kotera}
\author[tokushima]{Y.~Urano\fnref{kinken}}

%Kyoto 
\author[rigaku]{A.~Ichikawa}
%\author[kyoto]{K.Z.~Nakamura}
%\author[kyoto]{M.~Yoshida}

% LBL and UC Berkeley
%\author{T.I.~Banks}\lbl
%\author{S.J.~Freedman}\ipmu\lbl
\author[lbl]{B.~E.~Berger}
\author[lbl,ipmu]{B.~K.~Fujikawa}
%\author{K.~Han}\lbl ?????

%Hawaii
\author[hawaii]{J.~G.~Learned}
\author[hawaii]{J.~Maricic}

%MIT
\author[MIT]{S.~N.~Axani}
\author[MIT]{Z.~Fu}
\author[MIT]{J.~Smolsky}
\author[MIT]{L.~A.~Winslow}

% UT
\author[ut,ipmu]{Y.~Efremenko}

% TUNL
\author[tunl]{H.~J.~Karwowski}
\author[tunl]{D.~M.~Markoff}
\author[tunl,ipmu]{W.~Tornow}

% Virginia
\author[virginia]{S.~Dell'Oro}
\author[virginia]{T.~O'Donnell}

% Washington
\author[washington,ipmu]{J.~A.~Detwiler}
\author[washington,ipmu]{S.~Enomoto}

% NIKHEF
\author[nikhef,ipmu]{M.~P.~Decowski}
\author[nikhef]{K.~M.~Weerman}

%Boston U
\author[boston]{C.~Grant}
\author[boston,tunl]{A.~Li}
\author[boston]{and H.~Song}

\title{Search for Charged Excited States of Dark Matter with KamLAND-Zen}

\begin{abstract}
%% Text of abstract
Particle dark matter could belong to a multiplet that includes an electrically charged state.
WIMP dark matter ($\chi^{0}$) accompanied by a negatively charged excited state ($\chi^{-}$) with a small mass difference \mbox{({\it e.g.} $<$ 20 MeV)} can form a bound-state with a nucleus such as xenon.
This bound-state formation is rare and the released energy is \mbox{$\mathcal{O}(1-10$) MeV} depending on the nucleus, making large liquid scintillator detectors suitable for detection.
We searched for bound-state formation events with xenon in two experimental phases of the KamLAND-Zen experiment, a xenon-doped liquid scintillator detector.
No statistically significant events were observed. 
For a benchmark parameter set of WIMP mass $m_{\chi^{0}} = 1$ TeV and mass difference $\Delta m = 17$ MeV, we set the most stringent upper limits on the recombination cross section times velocity $\langle\sigma v\rangle$ and the decay-width of $\chi^{-}$ to $9.2 \times 10^{-30}$ ${\rm cm^3/s}$ and $8.7 \times 10^{-14}$ GeV, respectively at 90\% confidence level.
\end{abstract}

%%Graphical abstract
%\begin{graphicalabstract}
%\includegraphics{grabs}
%\end{graphicalabstract}

%%Research highlights
%\begin{highlights}
%\item Research highlight 1
%\item Research highlight 2
%\end{highlights}

\begin{keyword}
%% keywords here, in the form: keyword \sep keyword
Dark matter \sep Organic liquid scintillator \sep Xenon 
%% PACS codes here, in the form: \PACS code \sep code
%\PACS 0000 \sep 1111
%% MSC codes here, in the form: \MSC code \sep code
%% or \MSC[2008] code \sep code (2000 is the default)
%\MSC 0000 \sep 1111
\end{keyword}

\end{frontmatter}

%\linenumbers
%% main text

%%%%%%%%%%%%%%%%%%%%%%%%%%%%%%%%%%%%%%%%%%%%%%%%%%%%%%%%%%%%%%%%%%%%%%%%%%%%%%%%%%%%%%%%%%%%%%%%%%%%%%%%%%%%%%%%%%%%%%%%%%%%%%%%%%%%%%%%%%%
\section{Introduction}\label{sec:intro}
The constituent of dark matter (DM) is one of the most pressing problems in particle physics and cosmology~\cite{DarkMatter}.
It is expected to consist of one or more new, beyond the Standard Model (SM) particles.
A well-motivated class of hypothetical particles that are DM candidates are weakly interacting massive particles (WIMPs).
WIMPs are electrically neutral and are stable relative to the age of the universe.
The standard WIMP scenario assumes thermal equilibrium between DM and SM particles in the early stages of the Universe. 
In this case, the cross section times velocity $\langle\sigma v\rangle$, explaining the current dark matter abundance is $3\times10^{-26}$ cm$^{3}$/s.
However, there are wide varieties of dark matter models, and thus a broad dark matter search is needed to cover the diversity.
One example is the co-annihilation scenario where the masses of dark matter particles are degenerate~\cite{three_exception}. 
In such situations, constraints on $\langle\sigma v\rangle$ are greatly relaxed.

In this paper, we assume the existence of a DM doublet consisting of a WIMP and a new electrically charged particle.
If the WIMP is separated from the charged excited state by a mass difference $\Delta m$, the new particle is able to bind to nuclei~\cite{PRL} with an expected binding energy of $\mathcal{O}(1-10$) MeV.
Detectors sensitive to this energy region can be used to search for these interactions. KamLAND-Zen is such a detector, it consists of a large volume liquid scintillator detector with several hundreds kg of xenon nuclei and the signal is the bound state reaction with xenon nuclei. 
KamLAND-Zen has previously presented such a dark matter search~\cite{hosokawa}.
However, in this analysis we increased the statistics by including the KamLAND-Zen 800 data set, and optimized data analysis and simulation to perform an analysis with significantly improved sensitivity.
%%%%%%%%%%%%%%%%%%%%%%%%%%%%%%%%%%%%%%%%%%%%%%%%%%%%%%%%%%%%%%%%%%%%%%%%%%%%%%%%%%%%%%%%%%%%%%%%%%%%%%%%%%%%%%%%%%%%%%%%%%%%%%%%%%%%%%%%%%%
\section{The KamLAND-Zen detector}\label{sec:KL-Zen}
The KamLAND detector, located 2700 m.w.e. underground in the Kamioka mine in Gifu prefecture in Japan was originally designed to be a multi-purpose neutrino detector sensitive to anti-neutrinos, such as reactor-neutrinos and geo-neutrinos~\cite{kamland, KLreactor, KLgeo, KLgeo2022}.
It consists of 1 kton of liquid scintillator (outer-LS) filled in a 13-m-diameter spherical balloon made of 135-${\mu}$m-thick transparent nylon film and ethylene-vinylalcohol copolymer (EVOH).
This balloon is located at the center of an 18-m-diameter spherical stainless steel tank with 1325 17-inch and 554 20-inch photomultiplier tubes (PMTs) detecting scintillation light.
This detector part is called the inner detector (ID).
The remainder of the volume between the balloon and the stainless steel tank is filled with non-scintillating mineral oil.
A cylindrical water Cherenkov outer detector (OD) is constructed outside of the ID to identify cosmic ray muons and to serves as the radiation shield from the surrounding rock.

\begin{figure}[H]
\centering      
\includegraphics[width=\linewidth]{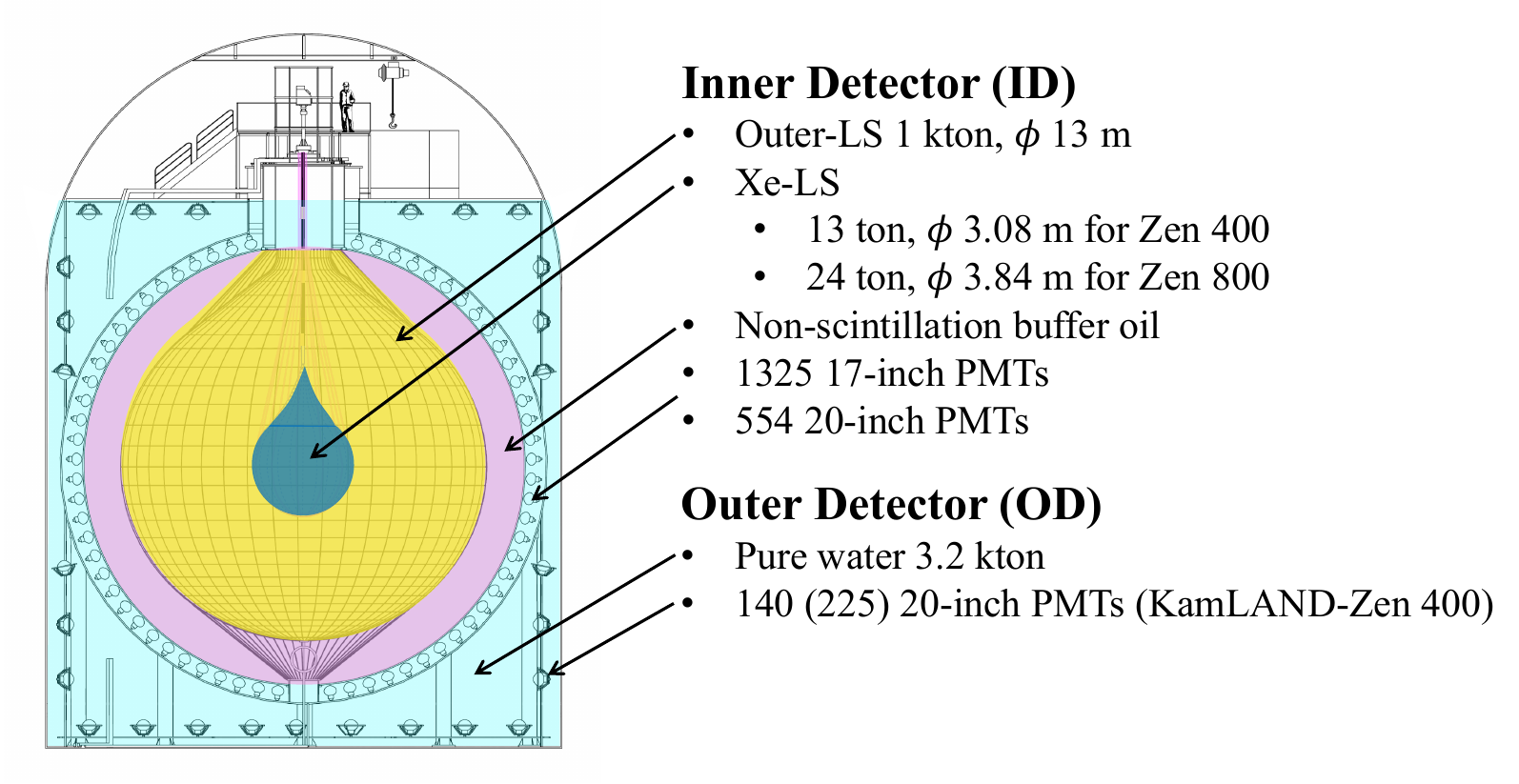}
\caption{\label{KamLAND_detector}
Schematic view of the KamLAND-Zen detector.
The IB radius is larger in KamLAND-Zen 800 than in KamLAND-Zen 400, doubling the amount of dissolved $^{136}$Xe.
} 
\end{figure}

KamLAND-Zen is a modification of the KamLAND detector.
The main goal of the experiment is to search for neutrinoless double-beta decays ($0\nu\beta\beta$) in $^{136}$Xe nuclei.
The first phase of KamLAND-Zen project was \mbox{KamLAND-Zen 400}, the second phase, \mbox{KamLAND-Zen 800}, is ongoing.
Fig.~\ref{KamLAND_detector} shows the KamLAND-Zen detector schematically.

\subsection{KamLAND-Zen 400}
\mbox{KamLAND-Zen 400 Phase-I} (2011-2012)~\cite{Zen400-1st} used 320 kg of xenon enriched with $^{136}$Xe nuclei.
The xenon-loaded liquid scintillator (Xe-LS) was in a 3.08-m-diameter, 25-$\mu$m-thick nylon spherical balloon (inner balloon, IB) located at the center of the KamLAND detector.
In this phase, $^{214}$Bi in the IB and $^{110m}$Ag were the main backgrounds.

\mbox{KamLAND-Zen 400 Phase-II} (2013-2015)~\cite{Zen400} was then started after purifying the Xe-LS.
This period was also divided into period-1 and period-2.
In this phase, $^{110m}$Ag was successfully reduced by purification.
The Xe-LS with about 381 kg of enriched xenon was filled in the same IB.
The Xe-LS was composed of 80.7\% decane and 19.3\% psedocumene (1,2,4-trimethylbenzene) by volume, 2.29 g/l of the fluor PPO (2,5-diphenyloxazole) and (2.91 $\pm$ 0.04)\% of enriched xenon gas by weight.
The isotopic abundances in the enriched xenon were measured by a residual gas analyzer to be (90.77$\pm$0.08)\% $^{136}$Xe and (8.96$\pm$0.02)\% $^{134}$Xe.
This is consistent with the component values of the quality specification sheet provided by the company.

\subsection{KamLAND-Zen 800}
\mbox{KamLAND-Zen 800} (2019-)~\cite{Zen800} has about 745 kg of enriched xenon by replacing the IB with a new 3.8 m in diameter balloon, doubling the amount of xenon compared to KamLAND-Zen 400.
This phase was successful in making a cleaner IB to reduce the BG events such as $^{214}$Bi~\cite{balloon}.

The Xe-LS is composed of 82\% decane and 18\% psedocumene by volume, 2.4 g/l of the fluor PPO and (3.13 $\pm$ 0.01)\% of enriched xenon gas by weight.
The isotropic abundance of the enriched xenon in this phase was given by the quality specification sheet provided by the company: $^{136}$Xe is (90.85$\pm$0.13)\% and $^{134}$Xe is (8.82$\pm$0.01)\%.
The uncertainty was estimated by the difference between the values in the quality specification sheets and the measured results in KamLAND-Zen 400.

The analysis used all xenon in the IB. 
The energy response of the Xe-LS and the outer-LS are different.
The relative light yield of the outer-LS to the Xe-LS is estimated to be 1.07~\cite{sayuri}.
This difference is taken into account in the present analysis.

%%%%%%%%%%%%%%%%%%%%%%%%%%%%%%%%%%%%%%%%%%%%%%%%%%%%%%%%%%%%%%%%%%%%%%%%%%%%%%%%%%%%%%%%%%%%%%%%%%%%%%%%%%%%%%%%%%%%%%%%%%%%%%%%%%%%%%%%%%%
\section{The models and the expected signal}
There are scenarios where a WIMP is part of a multiplet with an electrically charged excited state~\cite{PRL,Maxim2008,BaiFox2009,Nagata2015,Nagata2015PLB,AnnRev2020,Ellis2016}.
If the mass difference $\Delta m$ between the WIMP and the excited state is lower than \mbox{20 MeV}, the negatively charged excitation of the WIMP can form a stable bound state with xenon~\cite{PRL}.
The Coulomb binding energy, which depends on the target nucleus, compensates for $\Delta m$.

There are two possible cases: a positron is emitted (Case A) or a neutron in the nucleus is converted to a proton (Case B).
\begin{align}
&\mathrm{Case\,A:}\;N_Z + \chi^0 \rightarrow (N_Z\chi^-) + e^+,\\
&\mathrm{Case\,B:}\;N_Z + \chi^0 \rightarrow (N_{Z+1}\chi^-),
\end{align}
where $N_Z$ is the target nucleus with atomic number {\it Z} and $\chi^0$ ($\chi^-$) is the WIMP ground (excited) state.
The bound state $N\chi^-$ is formed in an excited state and will de-excite by emitting $\gamma$-rays.
In addition to the de-excitation $\gamma$-rays and a positron, annihilation $\gamma$-rays could be observed in this process for Case A.
In both cases, it is assumed that all expected signals are emitted simultaneously and the DM signal is detected as a single event in KamLAND-Zen.
Therefore, the total observable energy $E_\mathrm{tot}$ is given by~\cite{PRL},
\begin{equation}
E_\mathrm{tot}=\left\{\begin{array}{l}
\displaystyle E_b^{(0)}-\Delta m+m_e \;\;\; =K_{e^{+}}+E_{\gamma}+2m_{e}\;\;\;  \mathrm{( Case\,A )} \\
E_b^{(0)}-\Delta m+m_Z-m_{Z+1}\;\;\; = E_{\gamma}\;\;\;  \mathrm{ (Case\,B )},
\end{array} \right.
\end{equation}
where $\Delta m \equiv m_{\chi^-} - m_{\chi^0}$ and $E_{b}^{(0)}$ is the binding energy corresponding to the ground state of the bound states with the nucleus.
The Coulomb binding energy $E_{b}^{(0)}$, which depends on the target nucleus, has been calculated to be \mbox{18.4 MeV} for a xenon target~\cite{Maxim2008}.
Xenon nuclei are good targets because of their large $Z$ number, which increases the $E_{b}^{(0)}$ value and the $\Delta m$ region that can be explored.

\begin{figure}[H]
\centering      
\includegraphics[width=\linewidth]{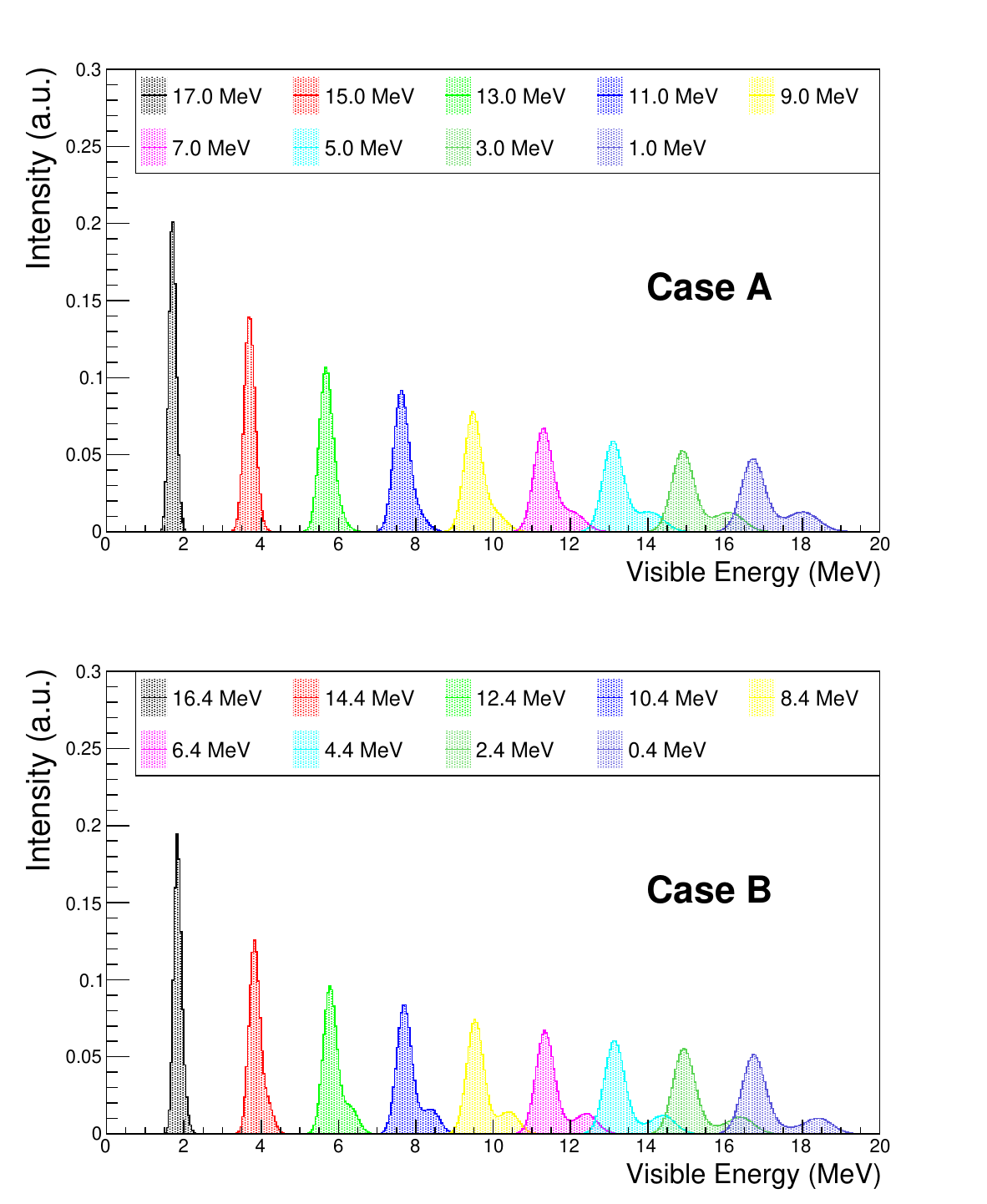}
\caption{\label{signal}
      The expected visible energy spectra of the bound state formation for several values of $\Delta m$ in the \mbox{KamLAND-Zen 800} detector.
       The multiple peaks at higher energy are due to the light yield differences between Xe-LS and the outer-LS.
      Top is for $\mathrm{Case\,A:}\;N_Z + \chi^0 \rightarrow (N_Z\chi^-) + e^+$ and bottom is for $\mathrm{Case\,B:}\;N_Z + \chi^0 \rightarrow (N_{Z+1}\chi^-)$.
} 
\end{figure}

Fig.~\ref{signal} shows the expected energy spectrum for several $\Delta m$ values, where the horizontal axis is the visible energy in KamLAND-Zen 800.
The higher and lower visible energy ranges correspond to lower and higher $\Delta m$, respectively.
These spectra take into account the detector response effects of the energy non-linearity, the energy resolution and the difference in light yield between the Xe-LS and the outer-LS as described in Refs.~\cite{Zen400, Zen800}.
The non-Gaussian distributions are due to the light yield difference in Xe-LS and outer-LS.
It depends on whether the emitted $\gamma$-ray deposits its energy mainly to the Xe-LS or the outer-LS.
In the low energy region, in contrast, almost all energy of the $\gamma$-ray is deposited in the Xe-LS, resulting in a single peak.

In the nuclear de-excitation, we assume that the energy corresponding to the difference of the binding energy is emitted as a single $\gamma$-ray with energy $E_{\gamma}$.
The expected number of signal events is given by Ref.~\cite{PRL}:
\begin{align}
  \label{eq:Nsignal}
  N_{\rm expected} = \frac{MTN_{\rm T}\rho_{\rm DM} \langle\sigma v\rangle}{2m_{\chi^0}},
\end{align}
where {\it MT}, $N_{T}$, $\rho_{\rm DM}$ and $\langle\sigma v\rangle$ are the detector exposure, number of target nuclei per kg, the local density of dark matter and the WIMP-nucleus recombination cross section with incoming dark matter velocity {\it v}, respectively.

Assuming a specific model as in Ref.~\cite{PRL}, the recombination cross section can be converted to other physical parameters.
In Case A, it can be converted to a decay width of a new particle $\Gamma_{\chi^{-}}$ with $\langle\sigma v\rangle$ as follows, 
  \begin{align}
    \label{eq:Bnl}
    \langle\sigma v\rangle &\simeq (|g_{eL}|^2 + |g_{eR}|^2) / (8\pi m_{\chi^{0}}) \times \sum_{n,l} B_{n,l},\\
    %B_{n,l} &\simeq (E_b^{(n,l)} - \Delta m - m_e) \sqrt{( E_b^{(n,l)} - \Delta m )^2 - m_e^2} \times
    %\int d^3r_1 d^3r_2 \phi^{\ast}_{n,l}(\vec{r}_1)\phi_{n,l}(\vec{r}_2)e^{i\mu\vec{v}\cdot(\vec{r}_1-\vec{r}_2)},\\
    \Gamma_{\chi^{-}} &= \tau_{\chi^{-}}^{-1} \simeq \frac{\sqrt{\Delta m^{2} - m_{e}^{2}}}{4\pi m_{\chi^{0}}} (\Delta m + m_{e}) (|g_{eL}|^2 + |g_{eR}|^2),
  \end{align}
where {\it $B_{n,l}$}, $\tau_{\chi^{-}}$ and $(|g_{eL}|^2 + |g_{eR}|^2)$ are the contributions from the capture into the state ({\it n,l}), the lifetime of $\chi^{-}$ and the Yukawa couplings, respectively~\cite{PRL}.
In Case B, assuming the Fermi gas model, the effective coupling $g_\mathrm{eff}$ is calculated using the following relations, 
 \begin{equation}
\begin{split}
\langle\sigma v\rangle&=\frac{g_\mathrm{eff}^4m_p^2}{8M_W^4}\int d^3r\rho_{n}(\vec{r})\int_{m_n}^{m_n+\frac{p_{nF}^2}{2m_n}}\frac{dp_n^0}{\frac{4}{3}\pi p_{pF}^3}\\
&\times\sqrt{(-V(r)-\Delta m+p_n^0)^2-m_p^2}\sqrt{p_n^{02}-m_n^2}\\
&\times\theta\left(-V(r)-\Delta m+p_n^0-m_p-\frac{p_{pF}^2(N_{Z+1})}{2m_p}\right),
\end{split}
\end{equation}
where $M_W$ and $V(r)$ are the $W$-boson mass and potential energy, respectively~\cite{PRL}.

%%%%%%%%%%%%%%%%%%%%%%%%%%%%%%%%%%%%%%%%%%%%%%%%%%%%%%%%%%%%%%%%%%%%%%%%%%%%%%%%%%%%%%%%%%%%%%%%%%%%%%%%%%%%%%%%%%%%%%%%%%%%%%%%%%%%%%%%%%%
\section{Analysis}\label{sec:anal}
In the present analysis, we used the \mbox{KamLAND-Zen 400} Phase-II period 1, 2 and \mbox{KamLAND-Zen 800} data-sets.

For the DM search, spherical fiducial volumes with 2.0 m and 2.5 m radius from the IB center were used for KamLAND-Zen 400 and KamLAND-Zen 800, respectively.
These selections cover the full volume of the IB and a small part of the outer-LS.
The reason for this selection is that the reconstructed position of the DM signal can be larger than the radius of the IB due to the emitted $\gamma$-rays.
The spatial detection efficiency $\epsilon _{\rm det} (r,{\rm \Delta m})$ is estimated from a Monte Carlo (MC) Geant4~\cite{g1,g2,g3} simulation. 
The spherical volume cuts have signal efficiencies of over 93\% and 94\% for \mbox{KamLAND-Zen 400} and \mbox{KamLAND-Zen 800}, respectively.
These cuts reject external backgrounds from the PMTs, the rock and the stainless-steel deck of the detector.

\subsection{Event selection}
Candidate events are selected by performing the following series of first level cuts, which are the same as in the $0\nu\beta\beta$ search~\cite{Zen800}.
\begin{enumerate}
\item
Events reconstructed within 0.7 m from the bottom hot spot on the IB are rejected in KamLAND-Zen 800 due to a relatively high background from $^{214}$Bi.
On the other hand, the KamLAND-Zen 400 IB has a 10 times higher background rate than the KamLAND-Zen 800 IB.
There is no equivalent cut in KamLAND-Zen 400 because it would cut away most of volume.
\item
  Muons and accompanying events within 2 ms after muons which cause PMT afterpulses and instability of the signal baseline, are rejected.
\item
  Sequential radioactive decays, such as $^{214}$Bi--Po and $^{212}$Bi--Po are tagged and rejected by a space-time correlation between the prompt and delayed events, called a delayed coincidence tag.
  The delayed coincidence tag requires that the time and distance between the prompt and delayed events be less than 1.9 ms and 1.7 m, respectively. 
  
 Because of the relatively short lifetime ($\tau$=430 ns) of $^{212}$Po, some $^{212}$Bi--Po decay sequences may be detected in a single event acquisition window. 
 These are tagged by looking for double-pulses in the PMT hit timing distribution.
 Due to the expected decay energy, the tag is only applied to events with visible energy of less than 5 MeV.
 The tag for identifying these pileup events is slightly different in KamLAND-Zen 400 and KamLAND-Zen 800 analyses. 
  In KamLAND-Zen 400, it does not lead to any inefficiency for the DM signal, while in KamLAND-Zen 800 it leads to a $\Delta m$ dependent inefficiency of a maximum of 6.3\% for $\Delta m$=17.8 MeV (Case A) and 9.5\% for $\Delta m$=17.2 MeV (Case B).
 For Case A, there is a possibility that the emitted positron will form ortho-positronium ($\tau$=3.2$\pm$0.1 ns in LS~\cite{ortho}). 
 We include a 2.9\% (0.6\%) systematic uncertainty in KamLAND-Zen 800 (KamLAND-Zen 400) to account for ortho-positronium.

\item
  Anti-neutrino events from reactors which produce a positron and a neutron from inverse ${\beta}$-decay are identified with the delayed coincidence tag.
\item
  Poorly reconstructed events with inconsistent charge, hit and time distributions are rejected to suppress electronic noise and accidental pileup.

\end{enumerate}

\subsection{Background}
The fiducial volume includes the region of the Xe-LS, the IB and the outer-LS.
One of the backgrounds for the DM search in both KamLAND-Zen 400 and KamLAND-Zen 800 are radioactive impurities in the fiducial volume. 
Other backgrounds are caused by neutrino interactions, external $\gamma$-rays or muon spallation products. 

\subsubsection{Radioactive impurities}
The primary background source arises from decays in the $^{238}$U and $^{232}$Th series in the Xe-LS, in the IB and outer-LS.
Other radioactive background sources are $^{85}$Kr in the Xe-LS and outer-LS, and $^{40}$K in the Xe-LS, outer-LS and IB film.
$^{110m}$Ag in the Xe-LS and IB is only considered a background in KamLAND-Zen 400 Phase-II~\cite{Zen400}.

\subsubsection{$^{136}$Xe two-neutrino double-beta decay}
 $^{136}$Xe two-neutrino double-beta decay ($2\nu\beta\beta$) has a Q-value of \mbox{2.458 MeV} and has an energy spectrum with a continuous distribution.
$2\nu\beta\beta$ events dominate in the energy region less than 2.5 MeV.

\subsubsection{Solar neutrinos}
Neutrino-electron elastic scattering events due to $^{8}$B solar neutrinos uniformly distribute in the detector.
The event rate is estimated to be (4.9 $\pm$ 0.2) $\times$ $10^{-3}$ (ton $\times$ day)$^{-1}$~\cite{Zen800}.
This energy spectrum continues up to about 18 MeV and is a primary background above 5.5 MeV. 
Additionally, solar neutrinos react with $^{136}$Xe through charged-current interactions producing $e^{-}$ and $^{136}$Cs ($\tau$=19.0 day, Q=2.548 MeV).
The decay of $^{136}$Cs to $^{136}$Ba gives a background peak at around 2 MeV.
The interaction rate is expected to be (0.8 $\pm$ 0.1) $\times 10^{-3}$ (ton $\times$ day)$^{-1}$ in Xe-LS~\cite{Zen800}.

\subsubsection{External ${\gamma}$-rays}
The origins of external ${\gamma}$-rays are the PMTs, the rock and the stainless-steel tank and the deck surrounding KamLAND. 
The 2.6 MeV ${\gamma}$-rays from $^{208}$Tl in the PMT's material are the dominant background. 
Additionally, ${\gamma}$-rays above 5 MeV are due to neutron capture in the rock and the stainless-steel structure.
%We fitted the radius distribution less than 5.0 m with attenuation length of ${\gamma}$-rays.
We fit the radius distribution for $r<5$ m with a function $f(x)\propto e^{(-x/\mu)}$ where $\mu$ is the attenuation length of ${\gamma}$-rays.
The estimated background rate in $r<2.5$ m for KamLAND-Zen 800 is (0.08$\pm$0.02)(day)$^{-1}$.
For KamLAND-Zen 400 the IB radius is smaller and the backgrounds on the IB are larger, making the external ${\gamma}$-ray background negligible. 

\subsubsection{Spallation}
Radioactive isotopes produced by cosmic muon spallation of carbon and xenon with accompanying neutrons, are uniformly distributed in the detector.
The decaying isotopes cause backgrounds for the complete $\Delta m$ of the DM signals under consideration.

These spallation events are selected with parameters based on time intervals from preceding muons ($\Delta T$), space correlations with vertices of neutron capture $\gamma$-rays induced by those muons, and reconstructed muon tracks and shower profiles~\cite{spallation}.
We divide the background into two groups according to their lifetime: short-lived and long-lived (LL), and apply the following four rejection criteria:
\begin{enumerate} 
\item Events within 150 ms after a muon passing through the LS are rejected. 
This cut removes 99.4\% of $^{12}$B ($\tau$ = 29.1 ms, Q = 13.4 MeV).

\item To reduce short-lived carbon spallation backgrounds, such as arising from $^{10}$C, $^{11}$C, $^{6}$He, $^{12}$B, $^{8}$B, $^{8}$Li and $^{11}$Be decays, we remove events reconstructed within 1.6 m from neutron vertices with $\Delta T$ \textless 180 s. 

\item To reduce the $^{137}$Xe ($\tau$ = 5.5 min, Q = 4.17 MeV) neutron capture background, we remove events reconstructed within 1.6 m from the vertices identified as neutron captures on $^{136}$Xe producing high energy $\gamma$’s (Q-value of 4.03 MeV) with $\Delta T$ \textless 27 min. 

\item 
Events without a neutron tag are rejected with the help of muon showers, by identifying the point around the spallation reaction as the largest energy deposition along the muon track~\cite{spallation}.
To reject the spallation products, we calculate the likelihood using probability density functions (PDFs) as follows, 
\begin{equation}
L_{1} = \mathrm{PDF}(\Delta Q,\,\Delta L) {\times} \mathrm{PDF}(\Delta T),
\end{equation}
where $\Delta L$ is the distance between the muon track and spallation background, $\Delta Q$ is the residual charge whose intensity implies the occurrence of energetic interactions in the detector and $\Delta T$ is the time interval from a preceding muon. 
The PDF($\Delta Q$, $\Delta L$) is made from $^{12}$B neutron-untagged events due to its short life time and high abundance. 
The PDF($\Delta Q$, $\Delta L$) of accidental events is estimated from an off-time-window analysis.
The PDF($\Delta T$) is evaluated using the Evaluated Nuclear Structure Data File (ENSDF)~\cite{ENSDF}

\item 
Some $^{136}$Xe spallation backgrounds with long lifetime (typically several hours) remain after these cuts.
The xenon spallation background is associated with more neutron production than carbon spallation because xenon is a neutron-rich nucleus.
To reduce the long-lived spallation background, we calculate another likelihood defined using neutron multiplicity as an ``effective number of neutron (ENN)'',
\begin{equation} 
L_{2} = \mathrm{PDF(ENN,}\,\Delta R_{nearest}){\times} \mathrm{PDF}(\Delta T),
\end{equation}
where $\Delta R_{nearest}$ is the distance from spallation backgrounds to the nearest neutron. 
We estimated the expected production rate using muon-induced spallation by FLUKA~\cite{f1, f2} with subsequent radioactive decays estimated by Geant4~\cite{g1}.
The systematic uncertainties of the production rate and the tagging efficiency for the likelihood method are also estimated by the MCs.
This tagging method is called LL-tag, and the LL-tagged events are also used in DM search fitting for long-lived spallation background estimation.
The tagging efficiency is determined in the spectral fit by introducing it as one of the fitting parameters.
The tag efficiency obtained by energy spectral fitting, which will be discussed in the next DM search section, are consistent with those estimated using MC.

\end{enumerate}

%%%%%%%%%%%%%%%%%%%%%%%%%%%%%%%%%%%%%%%%%%%%%%%%%%%%%%%%%%%%%%%%%%%
% Known background subtraction

\subsection{DM search}
We used the \mbox{KamLAND-Zen 400} Phase-II (period-1 and period-2)~\cite{Zen400} and \mbox{KamLAND-Zen 800}~\cite{Zen800} LL-untagged data-sets, with livetimes of 226.3 days, 221.8 days and 523.4 days, and  LL-tagged datasets with livetime of 16.6 days, 15.3 days and 49.3 days, respectively.
We searched for DM events in 0.5 MeV intervals in the $\Delta m$ range of \mbox{1$-$17.0 MeV} in the visible energy range of 0.8--20 MeV in KamLAND-Zen 400 and \mbox{1$-$17.0 MeV and 17.8 MeV} in the visible energy range of 0.5--20 MeV in KamLAND-Zen 800.

\begin{figure}[H]
\centering      
\includegraphics[scale=0.5]
{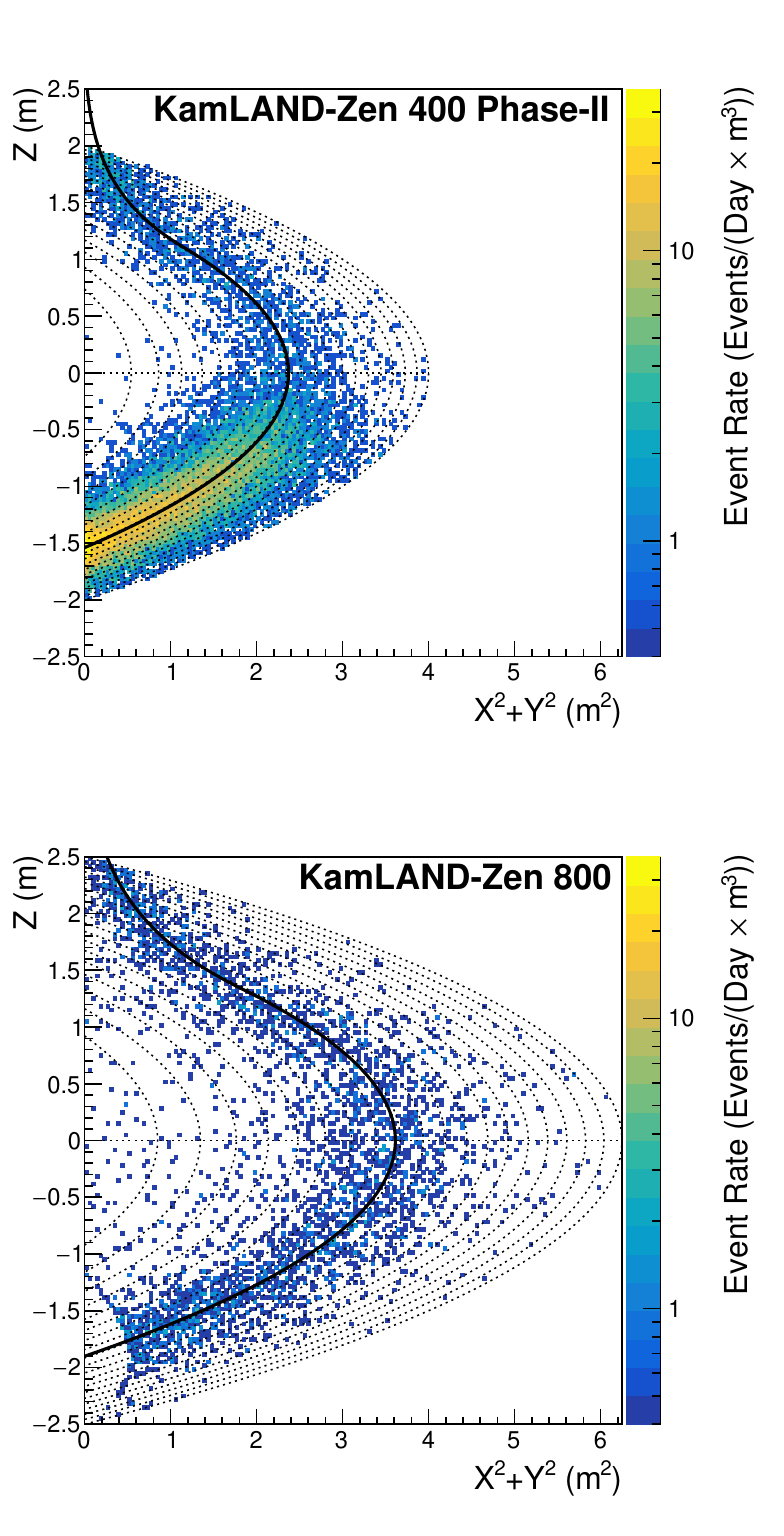}
\caption{
\label{vertex}
    (Top)The event rate distribution of candidate events in 3.0$-$20.0 MeV, above the $^{110m}$Ag background, in KamLAND-Zen 400 Phase-II.
    The thick black line is the IB film and the thin black dotted lines indicate the forty equal-volume spherical half-shells. 
    (Bottom)The vertex distribution of candidate events in the same energy region for KamLAND-Zen 800.
    The small region in the bottom part of IB is rejected from the analysis due to a relatively high background from $^{214}$Bi in the region.
}
\end{figure}

The DM rate is evaluated with the maximum likelihood method to the binned energy spectrum in concentric hemispherical-shell volumes, made by dividing the fiducial volume into 20 equal-volume bins for each of the upper and lower hemispheres, as illustrated in Fig.~\ref{vertex}.
The energy range of this figure is above the Q-value of 3 MeV for $^{110m}$Ag.

The energy region is divided into two categories because of the distribution of background: $\Delta m \leq 12$  MeV and $\Delta m > 12$ MeV.
%The region above 5.5 MeV has a low background due to low radiation from {\it e.g.} $^{238}$U and $^{232}$Th.
The region less than 5.5 MeV, the influence of IB-derived BG ({\it e.g.} $^{238}$U and $^{232}$Th) is significant, resulting in a complex BG distribution.
On the other hand, the major BGs in the region above 5.5 MeV are uniformly distributed in the Xe-LS and outer-LS, respectively. 
Therefore, the different number of volume-binning is employed.
We choose the following two situations for our fit:
\begin{description}
\item[For $\Delta m$ above 12 MeV]\mbox{}\\
In KamLAND-Zen 800, we fit to the binned spectra between 0.5 and 7.0 MeV in 0.05 MeV bins and divide the fiducial volume into 20 equal-volume bins for each of the upper and lower hemispheres as indicated in the lower panel of Fig.~\ref{vertex}.
The contributions from dominant backgrounds in the Xe-LS, IB and outer-LS, such as $^{85}$Kr, $^{40}$K, $^{210}$Bi, the $^{228}$Th-$^{208}$Pb sub-chain of the $^{232}$Th series, long-lived spallation products, and $2\nu\beta\beta$ are free parameters in the spectral fit.
The contributions from the $^{222}$Rn-$^{210}$Pb sub-chain of the $^{238}$U series and short-lived spallation products are constrained by their independent measurements~\cite{spallation}. \\
As for KamLAND-Zen 400 analysis, there are two points different from KamLAND-Zen 800: 
The first point is the addition of $^{110m}$Ag in the fitting as a free parameter, as already mentioned.
The other is that the lower edge of the fitting range is set to 0.8 MeV.

\item[For $\Delta m$ less than 12 MeV]\mbox{}\\
We fit to the binned spectra between 5.5 and 20.0 MeV in 0.5 MeV bins, and divide the fiducial volume into two equal-volume bins for each of the upper and lower hemispheres in KamLAND-Zen 800.
The contributions from short-lived spallation events such as $^{12}$B, $^{8}$Li, $^{8}$B and $^{11}$Be, solar neutrino elastic scattering and external $\gamma$-rays are constrained by their independent measurements.\\
In contrast, KamLAND-Zen 400 is not divided into upper and lower bins.

\end{description}

\begin{figure}[t]
\centering      
\includegraphics[width=\linewidth]{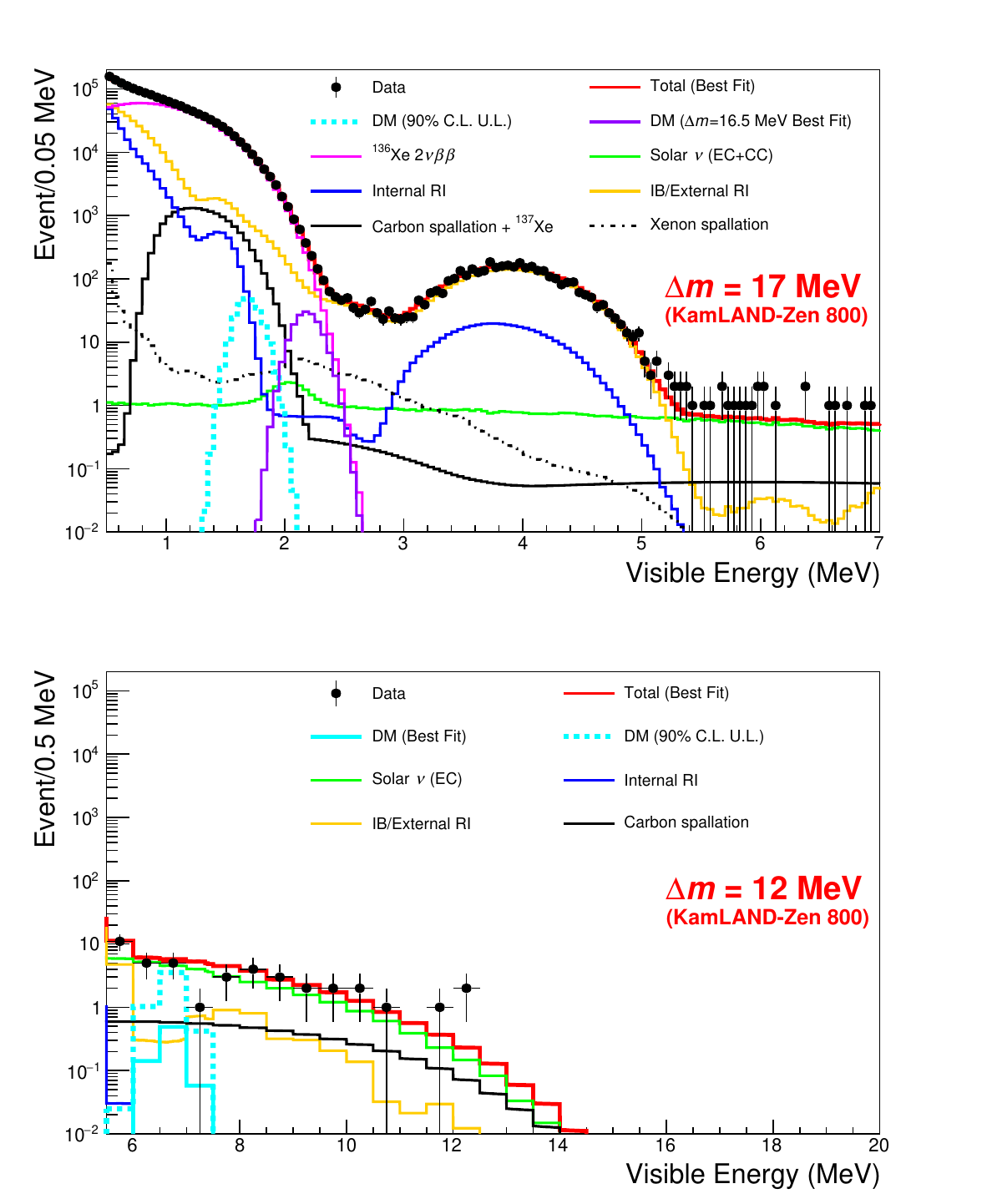}
\caption{\label{fit_result}
         Energy spectra of DM signal candidates within 2.5 m radius spherical volume with the best fit backgrounds, the best fit DM and 90\% C.L. upper limit.
         Top is for a DM signal of $\Delta m$=17.0 MeV and bottom is for a DM signal of $\Delta m$=12.0 MeV in \mbox{KamLAND-Zen 800} (Case A).
         At $\Delta m$=17.0 MeV, the DM rate of the best fit is 0.
         The results of the best fit at $\Delta m$=16.5 MeV are also overlaid in purple.
} 
\end{figure}

The best fit results for $\Delta m=17$ MeV and 12 MeV in \mbox{KamLAND-Zen 800} are shown in Fig.~\ref{fit_result}.
The cyan solid line is the best fit spectrum and the cyan dashed-line is the 90\% confidence level (C.L.) upper limit for DM.

\begin{figure}[t]
\centering      
\includegraphics[width=\linewidth]{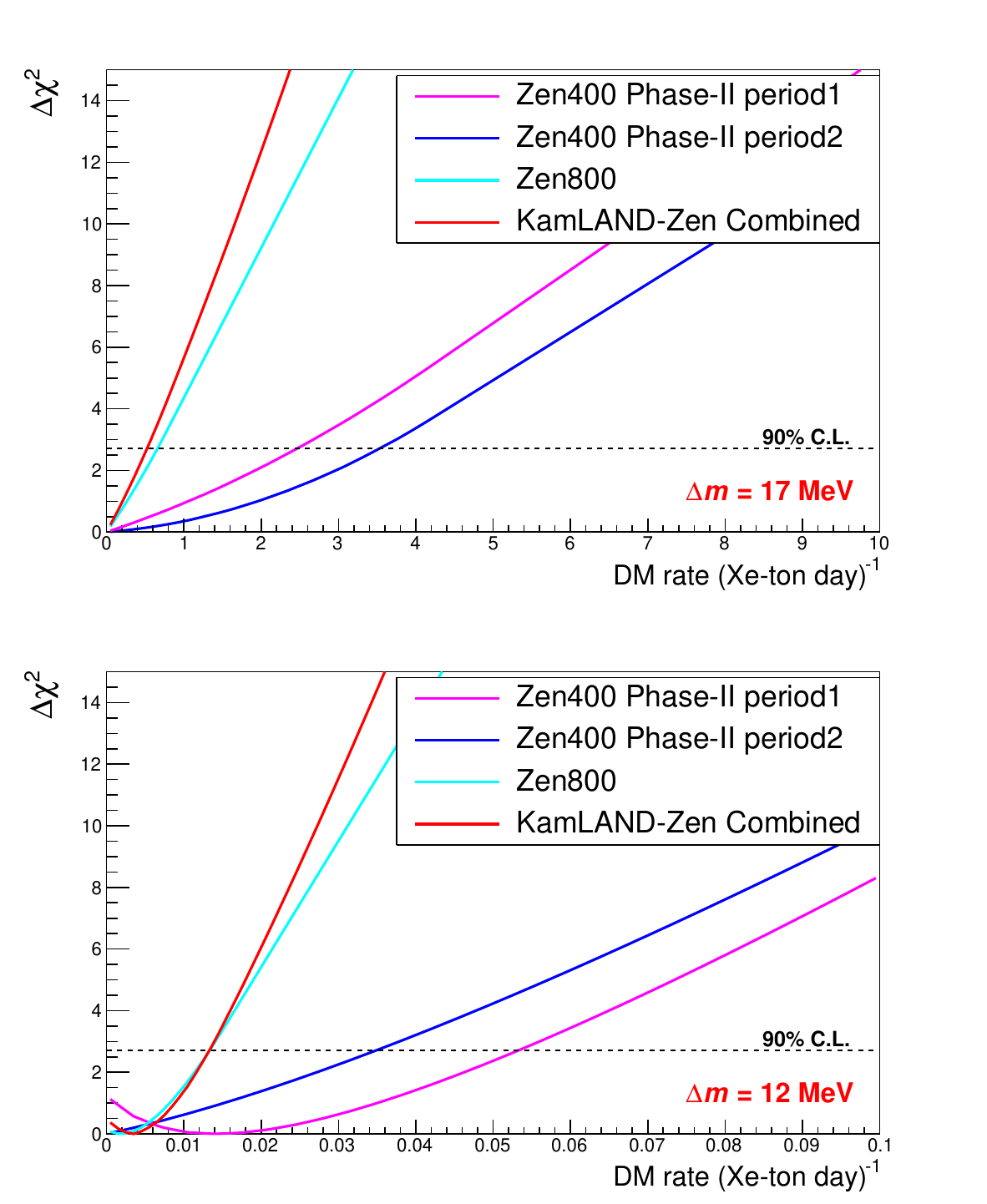}
\caption{\label{chi2_scan}
   The $\Delta\chi^2$ scan result for a combination of KamLAND-Zen 400 and 800.
   The horizontal axis is the event rate per xenon mass. 
   Top is for a DM signal of $\Delta m$=17.0 MeV and bottom is for a DM signal of $\Delta m$=12.0 MeV (Case A).
} 
\end{figure}

The results of the DM rate $\Delta\chi^2$ scan for KamLAND-Zen 400 and KamLAND-Zen 800 are combined in Fig.~\ref{chi2_scan} where the number of events per unit xenon mass is shown on the horizontal axis, in order to represent KamLAND-Zen 400 and 800 in one figure.

%%%%%%%%%%%%%%%%%%%%%%%%%%%%%%%%%%%%%%%%%%%%%%%%%%%%%%%%%%%%%%%%%%%
\section{Results}
Our search has a 3.4$\sigma$ local significance for $\Delta m=16.5$ MeV in KamLAND-Zen 800, which reduces to a 2.2$\sigma$ global significance when accounting for the look-elsewhere effect~\cite{Gross}.

The global p-value $p_\mathrm{global}$ is approximately the following with p-local $p_\mathrm{local}$,
\begin{equation}
p_\mathrm{global}\approx p_\mathrm{local}+\langle N_{u} \rangle,
\end{equation} 
where $\langle N_{u} \rangle$ is following,
\begin{equation}
\langle N_{u} \rangle \approx \langle N_{u_{0}} \rangle e^{-(u-u_{0})/2}
\end{equation} 

In order to calculate $\langle N_{u} \rangle$ with $u$=3.4$^{2}$ ,we took the following steps,
\begin{enumerate}
\item
Based on the background model, 16 data sets are generated with the same number of events as the data by MC.
\item 
The fitting in the $\Delta m$ range as in the DM search is performed and the significance of $\Delta m$ is calculated for each data sets.
\item 
$\langle N_{u} \rangle$ is calculated as the average number of up-crossings above the threshold $\sqrt{u_{0}}$=0.1.
\end{enumerate}

\begin{figure}[]
  \begin{center}
    \includegraphics[width=\linewidth]{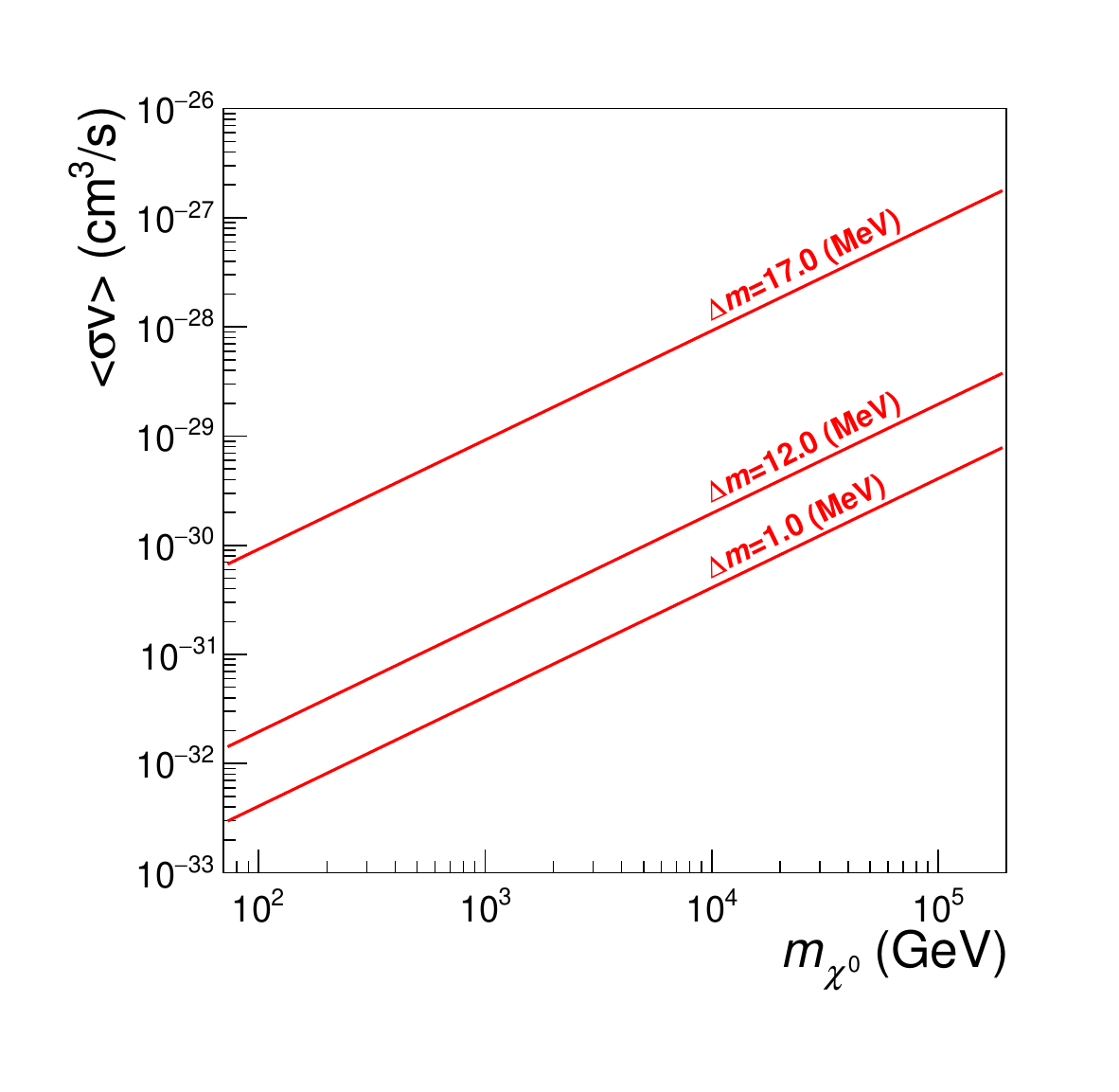}
    \caption{\label{resultsigmav}
      The WIMP-nucleus recombination cross section $\langle\sigma v\rangle$ as a function of the WIMP mass $m_{\chi^{0}}$.
      The red solid lines show the 90\% C.L. upper limit from this study for several $\Delta m$ values of KamLAND-Zen combined (Case A).
    }
  \end{center}
\end{figure}

\begin{figure}[]
  \begin{center}
    \includegraphics[width=\linewidth]{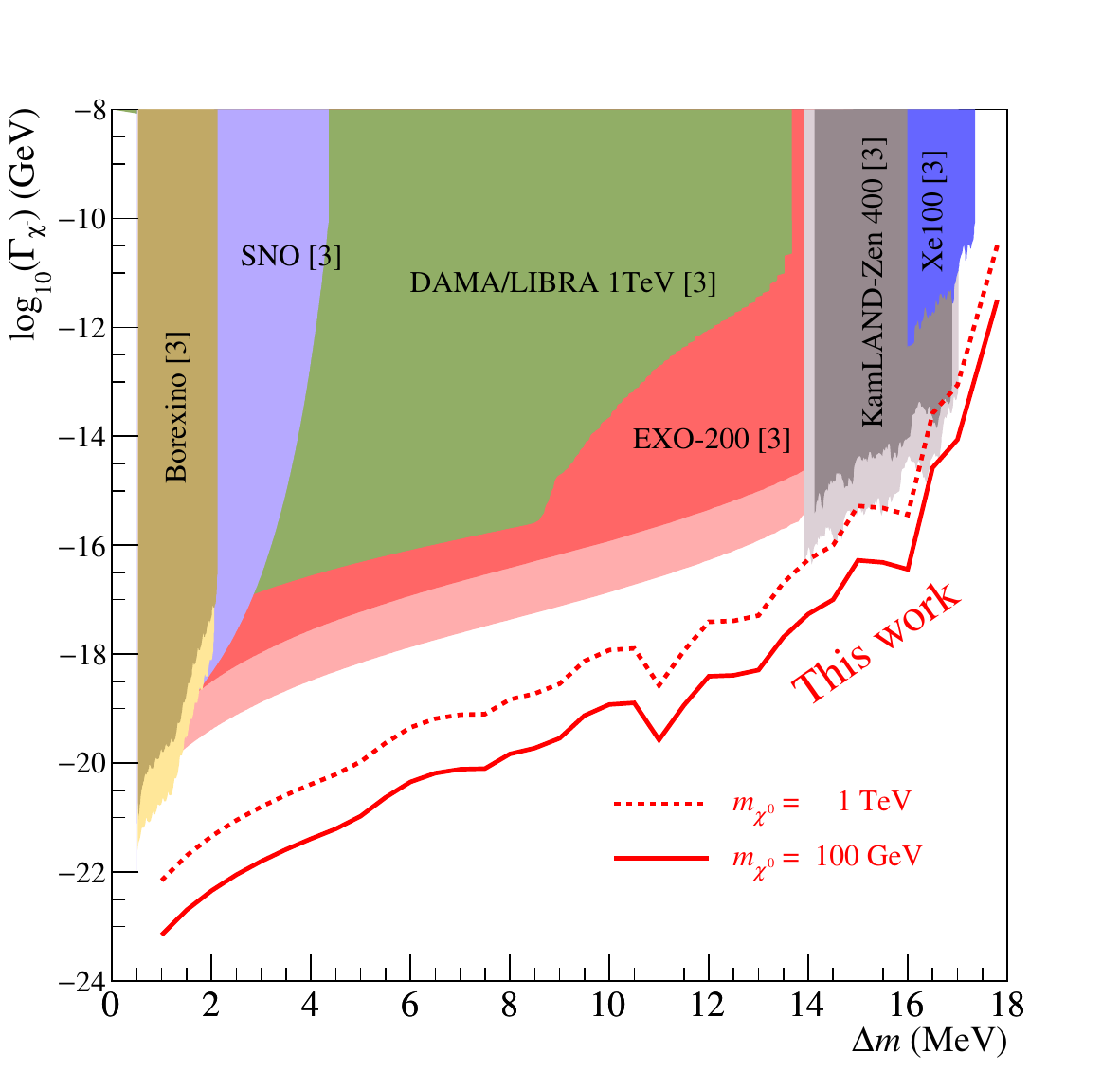}
    \caption{\label{result}
    Excluded decay width of $\chi^{-}$ as a function of $\Delta m$ for Case A.
      The red curves show 90\% C.L. upper limits of the KamLAND-Zen combined analysis.
The solid and the dotted lines correspond to $m_{\chi^{0}}$ of 100 GeV and 1 TeV, respectively.
      Filled regions are theoretical constraints using reported spectra from several experiments~\cite{PRL}.
      The darker and lighter regions correspond to constraints for 1 TeV and 100 GeV, respectively.
    }
  \end{center}
\end{figure}

\begin{figure}[t]
  \begin{center}
    \includegraphics[width=\linewidth]{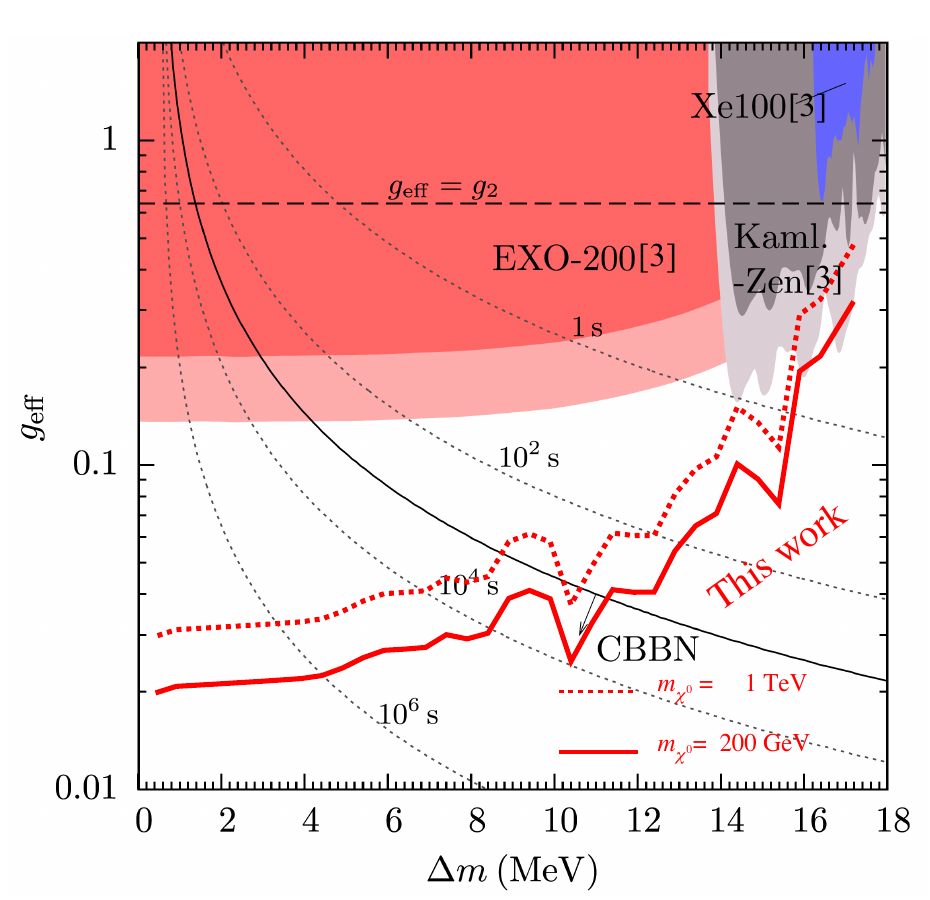}
    \caption{\label{resultB}
    Excluded effective coupling $g_\mathrm{eff}$ of $\chi^{-}$ as a function of $\Delta m$ for Case B.
      The red curves show 90\% C.L. upper limits of the KamLAND-Zen combined analysis.
      The solid and the dotted lines correspond to $m_{\chi^{0}}$ of 200 GeV and 1 TeV, respectively.
      Filled regions are theoretical constraints using reported spectra from several experiments~\cite{PRL}.
       The darker and lighter regions correspond to constraints for 1 TeV and 200 GeV, respectively.
      The catalyzed big bang nucleosynthesis(CBBN) shown by solid and dotted black lines was reported in Ref.~\cite{PRL}.
        }
  \end{center}
\end{figure}

No significant excess of events attributable to DM were found and we set upper limits at 90\% confidence level.
The results of this study for the WIMP parameter space are shown in Fig.~\ref{resultsigmav} to \ref{resultB}.

We  placed upper limits for the WIMP-nucleus recombination cross section $\langle\sigma v\rangle$ as a function of $m_{\chi^0}$ for several $\Delta m$ values, see Fig.~\ref{resultsigmav}.
The red curves in Fig.~\ref{result} and Fig.~\ref{resultB} show the upper limits on the $\Gamma_{\chi^{-}}$ (Case A) and $g_\mathrm{eff}$ (Case B), respectively, set by KamLAND-Zen combined.
Constrains based on reported spectra of several experiments~\cite{PRL} are also shown.
We obtain the most stringent limits on the charged excited states of WIMPs in the range of $\Delta m$=1.0$-$17.8 MeV (Case A) and $\Delta m$=0.4$-$17.2 MeV (Case B).

%%%%%%%%%%%%%%%%%%%%%%%%%%%%%%%%%%%%%%%%%%%%%%%%%%%%%%%%%%%%%%%%%%%%%%%%%%%%%%%%%%%%%%%%%%%%%%%%%%%%%%%%%%%%%%%%%%%%%%%%%%%%%%%%%%%%%%%%%%%
\section{Conclusion}
In summary, we performed searches for the bound state formation of a xenon nucleus and the electrically charged WIMP state using data from the $0\nu\beta\beta$ detector KamLAND-Zen.
No signals are found.
We set stringent limits: the recombination cross section upper limit for $\Delta m$=17.0 MeV is $\langle\sigma v\rangle < 9.2\times10^{-30}$ cm$^{3}$/s for $m_{\chi}$=1 TeV.
The upper limit on the new particle decay width for $\Delta m$=17.0 MeV is $\Gamma< 8.7\times10^{-14}$ GeV for $m_{\chi}$= 1 TeV.
This is corresponding to the upper limit of $g_\mathrm{eff}$ for $\Delta m$=16.4 MeV as $g_\mathrm{eff}< 0.32$ for $m_{\chi}$=1 TeV.
We plan to improve the rejection efficiency for the long-lived spallation background, continue to collect data and use an unbinned profiled likelihood to achieve a better sensitivity in the future.

\section*{Acknowledgments}
The \mbox{KamLAND-Zen} experiment is supported by JSPS KAKENHI Grant Numbers 21000001, 26104002, 19H05803 and 22H04934; the Dutch Research Council (NWO); and under the U.S. Department of Energy (DOE) Grant No.\,DE-AC02-05CH11231, as well as other DOE and NSF grants to individual institutions.
The Kamioka Mining and Smelting Company has provided service for activities in the mine. 
We acknowledge the support of NII for SINET4.

%% If you have bibdatabase file and want bibtex to generate the
%% bibitems, please use
%%

\bibliographystyle{elsarticle-num} 
%\bibliography{cas-refs}
\bibliography{reference}
\biboptions{sort&compress}
%% else use the following cocas-refs to input the bibitems directly in the
%% TeX file.

% \begin{thebibliography}{00}

% %% \bibitem{label}
% %% Text of bibliographic item

% \bibitem{}
% Create the reference section using BibTeX:

% \end{thebibliography}
\end{document}